\documentclass[12pt]{article}
\usepackage[utf8]{inputenc}
\usepackage[authoryear]{natbib}
\usepackage{xurl}
\usepackage[hidelinks]{hyperref}
\usepackage{graphicx}
\usepackage[margin=1in]{geometry}
\usepackage{wrapfig} 

\title{Genetic similarity versus genetic ancestry groups as sample descriptors in human genetics}
\author{Graham Coop\\
\small{Department of Evolution and Ecology and Center for Population Biology}\\ \small{University of California, Davis}\\
}
\date{July 2022}

\begin{document}

\maketitle
\begin{abstract}

A common sample descriptor in human genomics studies is that of ‘genetic ancestry group’, with terms such as ‘European genetic ancestry’ or ‘East Asian genetic ancestry’ frequently used in publications to describe the genetics of groups of individuals based on the analysis of their genotypes.  In this Perspective, I argue that these terms are imprecise and potentially misleading and that, for most applications, simple statements of genetic similarity represent a more accurate description.  
\end{abstract}

% https://www.nationalacademies.org/our-work/use-of-race-ethnicity-and-ancestry-as-population-descriptors-in-genomics-research
%I was asked to provide a commentary to the 
%\citeauthor{NAS_descriptors} “Use of Race, Ethnicity, and Ancestry as Population Descriptors in Genomics Research” committee. 

%This piece expands on my remarks from that talk. As a population geneticist, I focus on the use of genetic analysis to provide sample descriptors for human genetic samples, in particular the use of “genetic ancestry groups”. 
\section*{Introduction}

%Assigning sample descriptors is an initial step in the collection and use of medical and human genetics data. 
The sample labels researchers choose to apply to human genomic data shape downstream analyses and how results are interpreted by others. Some sample labels, such as race or ethnicity, are applied in ambiguous and inconsistent ways across studies  \citep{panofsky2017ambiguity,popejoy2020clinical,byeon2021evolving}. Today, sample descriptors also often include labels from the analysis of the samples’ genomic data. One common genetic sample descriptor that researchers use is `genetic ancestry group': for example, labelling individuals living in the United States as having `European genetic ancestry' or `African genetic ancestry'. 
 Because these labels are based on statistical methods, they may appear less arbitrary than labels assigned on the basis of social groupings.
However, the use of population genomics methods to assign
population descriptors comes with its own intersecting challenges. Indeed, genetic ancestry
labels are clearly a source of confusion, with slippage in scope and between genetic
versus social labels \citep[see ][for recent calls to be more precise about what we mean by genetic ancestry]{mathieson2020ancestry,lewis2022getting}.
 
 %Genetic sample descriptors are likely often unavoidable in practice, as scientists need them in order to combine and communicate findings.
 
In this perspective, I argue that the field of human genetics should move away from using
genetic ancestry groupings as sample descriptors. Such terms are imprecise and potentially misleading and, for most applications, researchers use them to indicate genetic similarity or relatedness to some predefined set of samples.  In most applications, human geneticists are actually concerned with controlling for genetic similarity, geography, and environments in their comparisons rather than some vague notion of ancestral populations. A number of subfields of human genetics, including human population genetics and genetic anthropology, are more directly concerned with understanding history by reconstructing various aspects of genetic ancestors; in my view, this is largely a related but distinct enterprise from providing the field of human genetics with useful genetic sample descriptors. 
 
 Given the issues associated with ancestry group labels, I believe that much of human genetics research should move towards using more readily interpretable statements about genetic similarity (and relatedness) for sample descriptions. These statements are often nearly equivalent in terms of the information they contain, but simple statements about genetic similarity/relatedness are a more accurate statement of what population-genetic methods are providing, and importantly such language carries far less baggage in terms of its implicit depiction of the structure of human groups.

I’ll first begin with a few words on human genetic variation before turning to why genetic sample descriptors will likely be necessary for some time to come. I’ll discuss the topic of genetic ancestors and why in practice genetic ancestry description labels used in human genetics boil down to genetic similarity. I’ll briefly touch on some of the issues and responses raised by genetic ancestry labels, before finishing with an outline of why genetic similarity labels are a better set of descriptors for the broad field of human genetics. 

\section*{Genetic variation}
Genetic populations are modelling constructs. As in other scientific endeavours, when developing mathematical models of evolution and statistical tools to analyse data, we have to simplify reality. In such models, we often assume that there are well-defined populations, which are modelled as persisting in particular areas, and that represent somewhat separate and well-defined evolutionary lineages \citep[e.g.][]{gutenkunst2009inferring,excoffier2013robust}.  Such models can be incredibly useful in helping shape our intuition and analysis. However, they also can constrain our thinking and lead to confusion. 

Well-delimited genetic populations are rarely found within species in the real world. Certainly, such populations are very rare in humans. We can of course talk of `populations' consisting of everyone in the UK, Finland, or whole continents such as Africa, but these are obviously not populations in the sense that is typically meant in evolutionary genetics, in which individuals within a group share closer genealogical relationships to each other than any other individual outside the group. Even if we take subsets of people within these regions---for example, people who self-identify as white in the UK or people whose four grandparents all come from the UK--- these groups will not form well-delimited genetic subsets, as the UK is not a homogeneous population with respect to migration in and out of the UK \citep{leslie2015fine,olalde2018beaker,patterson2022large}. With sparser sampling of a limited number of geographic locations, we can sometimes identify groups that are closer to better defined `genetic populations (e.g. the splits among the non-American 1000 genomes samples in Fig \ref{Fig:Martin_PCA}), but these simply reflect the limitations of our sampling.
 
The reality is that we are all related to each other to varying extents, in a complex web of genealogical relations that form an unimaginably complicated family tree. As a result, genetic variation varies fairly smoothly among individuals, often in ways that are correlated with environments. Patterns of human genetic variation are shaped by geographic distance, geographical barriers, as well as broad-scale population movements \citep{ramachandran2005support,wang2012quantitative,peter2020genetic}. It might be tempting to think that the fairly continuous nature of modern human variation reflects admixture only over the past few hundred years. However, human groups of the past have often been geographically widespread and ephemeral, frequently forming, only to rapidly collapse together with other nearby and sometimes much more distant groups. This is a point that advances in ancient DNA technology have repeatedly made abundantly clear \citep{skoglund2018ancient,liu2021insights}.

Much of the common genetic variation found within groups is shared across human groups  (\citealp{Lewontin1972}; \citealp[for a visualization see][]{biddanda2020variant}). Rarer genetic variation is usually more localized \citep{gravel2011demographic,nelson2012abundance}. In fact, individual rare variants are better thought of not as properties of groups that bear any resemblance to the ancestry groupings used in genetics research, but rather as features of extended families within such groups. Local adaptation is sometimes highlighted as a reason that particular phenotypic outcomes might be common in some groups and rare in others (i.e. due to combinations of alleles varying substantially in frequencies across groups). Although a growing number of convincing cases of genetic variants shaped by local adaptation have come to light \citep{fan2016going}, these are still a tiny minority of the many loci associated with phenotypic and disease outcomes. Furthermore, locally adapted alleles do not map onto ancestry group as selection pressures vary continuously and convergent adaptation among geographically separated groups is common \citep{ralph2010parallel,jeong2014adaptations}. 

As a consequence of the complicated genetic structure of humanity, there is no single “right” level of granularity to use for description for all questions. Human groups are structured from broad geographic scales to fine-scale patterns well below the level of a country \citep{leslie2015fine,han2017clustering,liu2018genomic,narasimhan2019formation,raveane2019population,bycroft2019patterns,byrne2020dutch}. The choice of level of granularity will depend on the specific questions being addressed. For example, the accuracy of polygenic score predictions is lower for people of Italian and Polish ancestry in the UK than for people of United Kingdom ancestry \citep[][the relative contributions of changes in linkage disequilibrium, specific genetic, environmental and gene-environment interactions to such changes in accuracy are under current investigation and debate]{Prive2022}. Whether the researcher thinks a set of polygenic scores are appropriate for use in `European ancestries' or that they need much more fine-grained GWAS will depend on the questions being pursued and the prediction resolution required by any future uses. Thus, the nature of the sample descriptions needed in this example, and in many other cases in human genetics, depends on the problem at hand. 

Given the complexity of human genetic variation, all verbal descriptions of genetic structure will be incomplete and open to miscommunication.  While this has often been an problem in human genetics, this issue is coming to the fore as the depth of genetic sampling of humans increases. 
Thus, we need to move towards terminology for genetic sample descriptors that is clear to people across fields and encourage good use. 

\section*{Descriptors or labels in analysis.}
If human genetic structure is fairly continuous, why then do we need to use (somewhat) discrete sample descriptors? Lots of valid explanations and uses have been put forward \citep[see for example talks from the][]{NAS_descriptors_1,NAS_descriptors_2,NAS_descriptors_3}. From my perspective, as a population geneticist, there are two major reasons for using genetic sample descriptors that I want to highlight. 

\subsection*{Data subsetting}
%Researchers set out to gather large enough samples that will offer sufficient statistical power to study their chosen problem. But in practice, the sampling, genotyping, phenotyping, and analysis effort of a given project will always be limited. 

There are obviously many historical and ongoing reasons why sampling has focused on particular groups, but one technical issue is the limitations of statistical approaches and tools.  Many methods in statistical and population genetics fit statistical models that assume relatively homogeneous groups. For example, standard GWAS rely on pseudo-randomization of genotypes at a locus across genomic backgrounds and environmental causes of trait variation \citep{vilhjalmsson2013nature}, and so researchers often limit themselves to more genetically well-mixed groups to better satisfy these requirements \citep[although issues of residual heterogeneity will remain, ][]{haworth2019apparent,zaidi2020demographic}. On the population genomics side, methods to reconstruct population history are constrained to describing the history of a small number of groups, while methods to describe more realistically continuous histories are daunting, and as yet underdeveloped. Because of these limitations, researchers subset their data in various ways, for example restricting GWAS samples to a particular ethnicity and geographic region. But having genotyped the individuals, they often further subset their data to a particular subset of genetic variation, for example, a particular subregion of principal components analysis (PCA) space where many of their samples are present \citep[e.g. the UK Biobank's "white British ancestry subset" ][]{bycroft2018uk}. Having subset the samples, descriptors are needed for these genetic subsets (“Who was the method applied to?”).   

Some of the limitations of methods reflect historical legacies of working in modelling frameworks designed to analyse smaller genetic datasets and previous computational limitations. Methods are improving, allowing GWAS to be performed across somewhat more heterogeneous samples and more flexible models of human population genetic history \citep{brown2016transethnic,novembre2016recent,peterson2019genome,patel2022genetic}.  However, conceptually, it is very hard to analyse even small subsets of human diversity all at once and so it is unlikely that subsetting genomic data will stop in practice any time soon. Thus, genetic sample descriptors will remain in use for the foreseeable future. 

\subsection*{Communication}
The second and perhaps bigger reason for genetic sample descriptors is that scientists often rely on them to communicate their findings to others.  In many human genomic analyses, assessing patterns of genetic structure is the first step. In such applications, researchers use prior population descriptors to orient results: “What do the observed patterns reflect?”. For example, what features do the major axes of variation found in a PCA correspond to? We need words to describe these axes and describe our interpretations. Furthermore, as a field we often pool together different datasets, for example, in GWAS meta-analysis, or combine together different data types, for example, GWAS effect sizes from one group with genotype data from another. In a practical sense, we need shorthand labels to communicate what we are doing.

\section*{Genetic ancestry}
 `Genetic ancestry' is an evocative concept, as made clear by the success of personal genomics companies, which capitalize on the fact that our views on ancestry and identity long predate the development of genetics. I first want to make the distinction between two related concepts of `genetic ancestors' and `genetic ancestry group'. The former is a well-defined concept and a central part of population genetics. 

\subsection*{Genetic ancestors}

Your genealogical ancestors, the people from whom you are biologically descended, are a well-defined set of people. More than a few hundred years back, each of us have tens of thousands of genealogical ancestors, a number that initially grows exponentially backward in time until it stabilizes when you are descended from everyone who left any descendants to the present day. However, with only two copies of your genome, which is inherited in big blocks, you cannot inherit genetic material from all of these ancestors. Instead, you inherit genetic material from only a very small subset of your ancestors \citep{donnelly1983probability,Coop_How_many_genetic_anc}. For example, $\sim$450 years ago you may have more than 32,000 living genealogical ancestors, but only $\sim$1,000 of them contributed genetic material that ended up in your genome, and the proportion of ancestors who contribute genetic material drops farther back in time. The subset of your genealogical ancestors from whom you inherited genetic material are your `genetic ancestors', a small fraction of your total ancestry.

Several thousand years back, all modern humans share all of their genealogical ancestors. Farther back than that, anyone who left any descendants in the present day (and many did) is an ancestor to all humans living today \citep{manrubia2003genealogy,rohde2004modelling,Coop_genealogical}. What does this mean for our genetic relatedness? Well, that’s complicated. You and I share a genealogical common ancestors at least as recently as the time when all modern humans share all their ancestors. However, only a limited subset of these people are our genetic ancestors, and we only inherit small fractions of our genome from any one of these common ancestors. Therefore, at a typical locus, your and my most recent genetic common ancestor lived much farther back in the past.

Genetic similarity, genealogy, and genetic ancestry are closely related concepts, in ways that are non-trivial to understand \citep{Coop_Genetic_ancs}. All four of my grandparents came from Northern England. As a result, I share slightly more genetic variants in common with other people whose grandparents all came from Northern Europe than I shared with people whose grandparents came from elsewhere in the world. That’s because, while everyone in the world shares all of their genealogical common ancestors several thousand years ago, they are not weighted the same in different people’s genetic family trees. A particular ancestor could appear tens or hundreds of times tracing back along different paths in my family tree if they are many generations back. That same ancestor many generations back could appear only once in someone else’s family tree with only a single path to them. While they are an ancestor to both of us, I’m somewhat more likely to have inherited a (small) chunk of my genome from them than the other person is. Even though we all share many genealogical ancestors, I have many more paths back through my family tree to ancestors who are also ancestors many times over for other Northern Europeans than I do with someone from, say, Japan. As a result, I share slightly more genetic variants in common with another Northern European than I do with a person whose grandparents all came from Japan. My genetic resemblance to many Northern Europeans reflects the fact that we share somewhat more of our genealogical ancestry --- but not in a way that maps simply onto statements that my ancestors are all European or that Europeans share some set of ancestors that people from elsewhere do not. 
\begin{wrapfigure}{r}{0.4\textwidth}
  \begin{center}
   \includegraphics[width=0.39\textwidth]{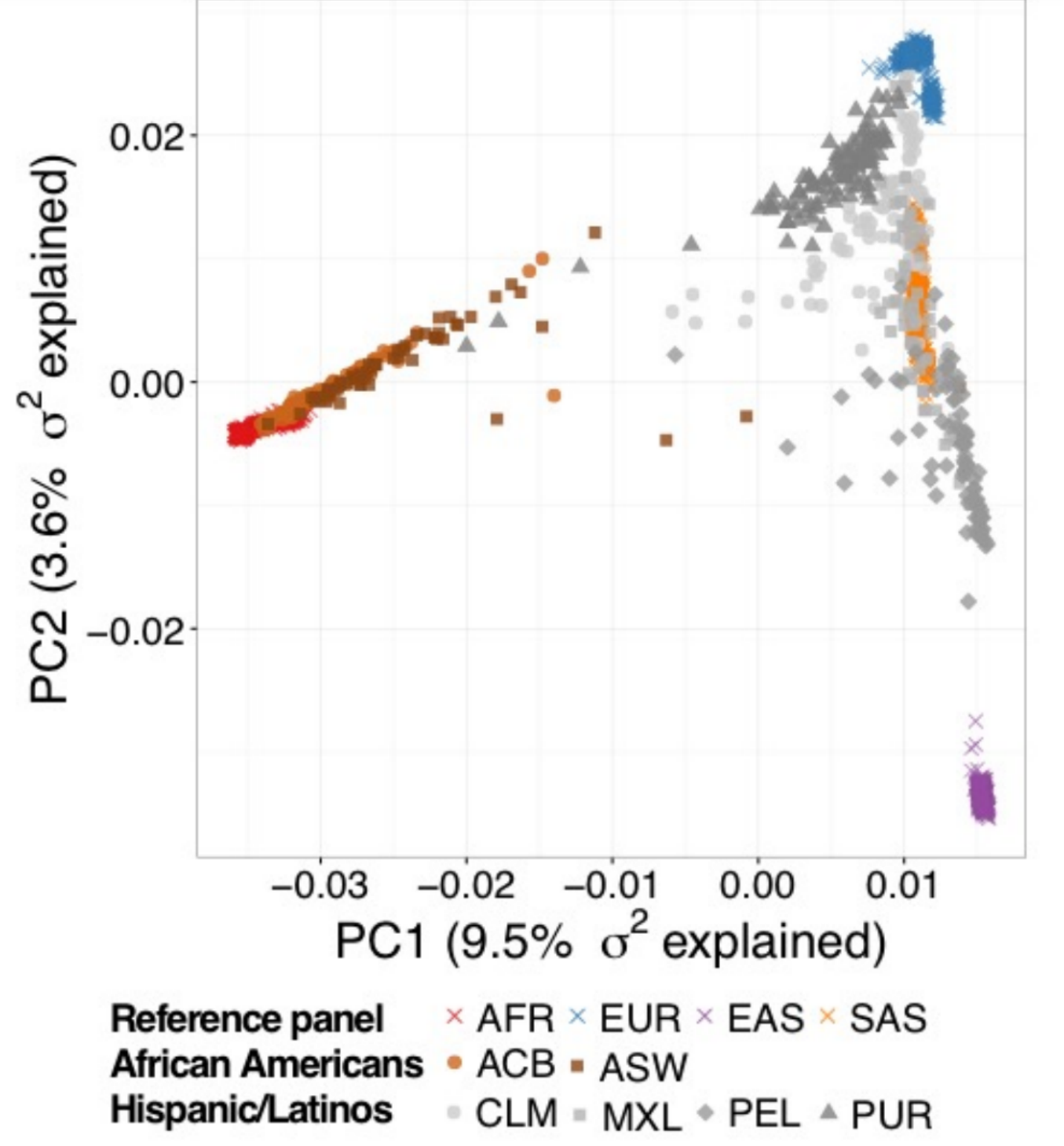}
  \end{center}
  \caption{{\small Figure from \citet{martin2016population} using 1000 genomes samples: ``Principal components analysis of all samples showing the relative homogeneity of AFR, EUR, EAS, and SAS continental groups and continental mixture of admixed samples from the Americas (ACB, ASW, CLM, MXL, PEL, and PUR)." Cropped from Figure 1 in preprint (CC BY-NC 4.0), figure S1 in published paper. 1000 genome sample codes given at this \href{https://uswest.ensembl.org/Help/Faq?id=532}{link}.}
} \label{Fig:Martin_PCA}
\end{wrapfigure} 
Our genetic ancestors are a well-defined set of people, who lived in particular places and times. Obviously, in practice, we do not know who they were, but we could hope to infer things about them. Population genomic data can inform us about our relatedness, our shared genetic common ancestors, through summaries of genetic similarity among individuals, which reflects the sharing of genetic material transmitted through meiosis \citep{rosenberg2002genealogical}. A lot of the statistical machinery of population genetics builds on these ideas to learn about evolutionary processes and history.

However, much of our interpretation of these patterns comes from combining this information with geographic and sample descriptors of the analysed genomic data. Thus, our interpretations and genetic sample labels always reflect, at least in part, the social context of how samples were chosen and described, and thus they are partially social constructs.  

Through computational and statistical advances, the field of population genetics is getting much better at describing some properties of our vast number of shared genetic ancestors. A major recent breakthrough is the development of approaches to computationally reconstruct the so-called `ancestral recombination graph', or approximations of it, for large genomic samples \citep{speidel2019method,kelleher2019inferring}. The ancestral recombination graph describes the full set of genetic relationships among a set of samples in terms of their shared genetic ancestors \citep{hudson1990gene,marjoram1995ancestral}.  Along with this breakthrough has come the hope that approaches building on the ancestral recombination graph will allow a fuller description of “genetic ancestry”. It is doubtlessly true that these advances will allow us a fuller picture of some of the properties of our vast clouds of genetic ancestors and underscore the point that we are all embedded in the same giant tree of humanity, something other methods can obscure.  However, these representations will necessarily often be high-dimensional and do not lend themselves easily to verbal summaries. 

\section*{Defining genetic ancestry groups}
When human genetic researchers use ancestry group terms such as “European ancestry” or “East Asian ancestry” as a sample descriptor they are (nearly) always a description of genetic similarity to other present day individuals by some summary statistics \citep{mathieson2020ancestry}.

\begin{wrapfigure}{r}{0.46\textwidth}
  \begin{center}
   \includegraphics[width=0.45\textwidth]{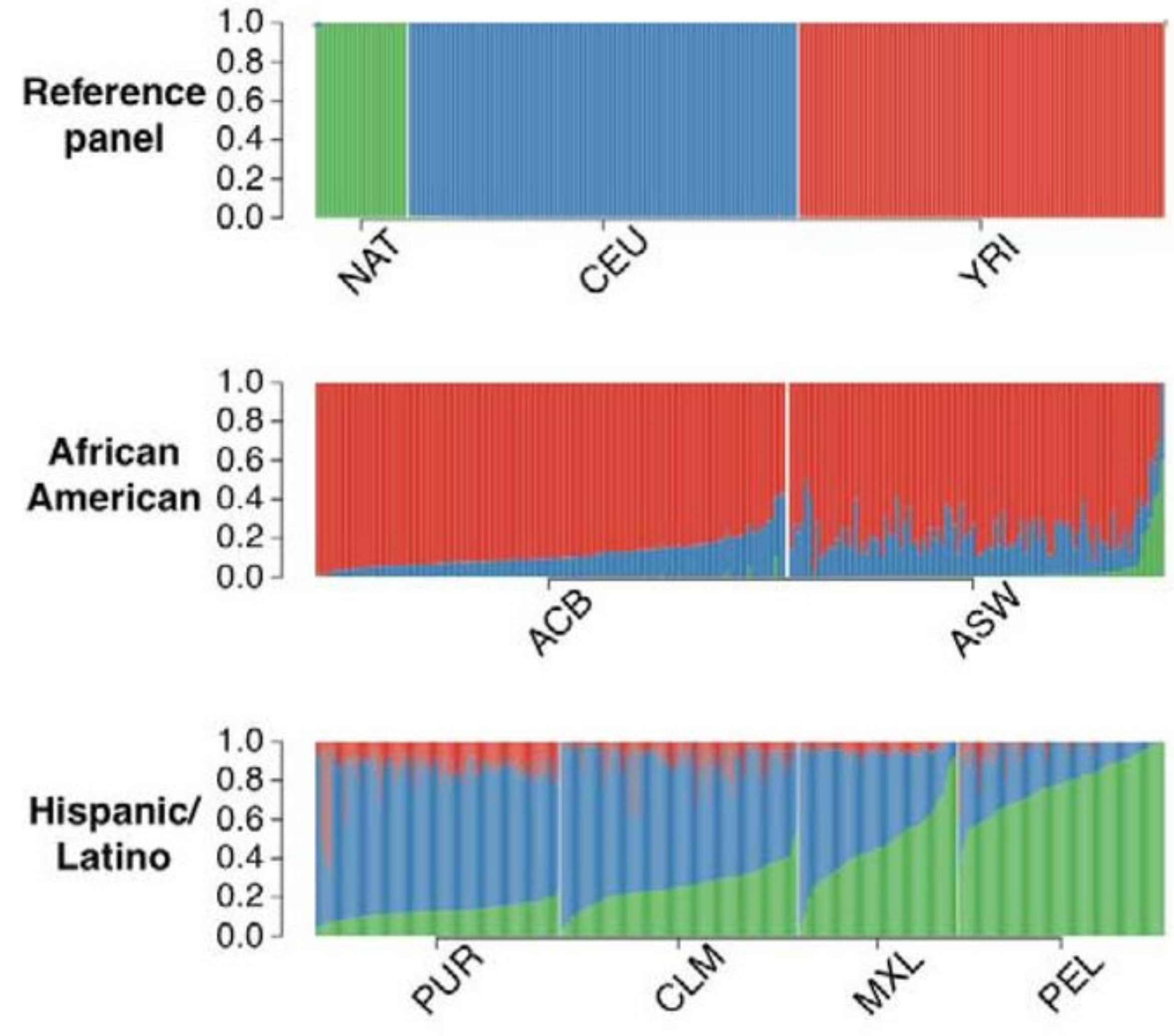}
  \end{center}
  \caption[-2]{{\small Figure from \citet{martin2016population} from their caption ``ADMIXTURE analysis at K = 3 focusing on admixed Americas samples, with the NAT, CEU, and YRI as reference populations.'' They find that ``[t]he six populations from the Americas demonstrate considerable continental admixture, with genetic ancestry primarily from Europe, Africa, and the Americas''. NAT is a sample of Native Americans from \citet{mao2007genomewide}. Figure cropped from Figure 1 of preprint (CC BY-NC 4.0). 
}} \label{Fig:Martin_admixture}
\end{wrapfigure} To illustrate this, we can work through some common ways in which ancestry groups are assigned in human genetics. In doing this, I note that I have certainly referred to these approaches in these terms in the recent past, and it can be a very convenient way to explain the concepts at play. My discussion of this topic is forward-looking rather than a judgment of past uses, including my own.  

One of the most common ways that ancestry labels are assigned to samples is on the basis of how a person’s genome clusters with other samples in a genetic PC plot. For example, Figure \ref{Fig:Martin_PCA} shows the 1000 Genomes samples positioned on their first two genetic principal components. If you projected my genome onto this plot, my genotype would doubtless cluster with the samples labelled ‘EUR’. On that basis, researchers might choose to label me as belonging to the `European ancestry group'. However, the fact that I would fall close to samples labelled `European' is simply a statement that my genotype is similar to the genotypes of those people along the axes of variation captured by the top two PCs. It is also broadly a statement that I shared a higher degree of relatedness with these individuals along these axes \citep{mcvean2009genealogical} but it is not a statement that my genetic ancestors form a delimited group with other such individuals.

%https://pan.ukbb.broadinstitute.org/docs/study-design#how-did-you-decide-what-ancestry-groups-to-include-how-did-you-assign-individuals-to-each-ancestry-group
Conversely, imagine a person who comes with a sample descriptor `English', for example, based on a self-identified label. If this person has six great-grandparents from England and two whose ancestors trace recently back to West Africa via the Caribbean, their genome might fall roughly three-forths of the way between the African and European 1000 Genomes panel samples on this plot. On that basis, researchers may choose to exclude them from the `European ancestry' group of individuals. Indeed, researchers usually predefine some set of cutoffs, for example, if a person falls more than 3 standard deviations from the centroid of the cluster of individuals labelled `European' individuals  then they are not retained in the European ancestry data subset. As discussed above, there can be good methodological reasons for wanting a relatively genetic homogeneous group for analysis. However, the decision of whom to include is necessarily a somewhat arbitrary exercise in applying discrete labels to continuous variation.  

%engelhardt2010analysis
Other methods of assigning ancestry groups allow for a person’s ancestry to be drawn from multiple “ancestral” groups \citep[{\tt STRUCTURE} and {\tt ADMIXTURE} style approaches:][]{pritchard2000inference,falush2003inference,alexander2009fast}; see Figure \ref{Fig:Martin_admixture} for an example. These methods would allow a breakdown, for example, of someone’s ancestry into 75\% European and 25\% African ancestry. In this case, alleles are modelled as being drawn from some hypothetical well-mixed populations; however, the `genetic ancestry' labels for those populations are being propagated by investigators from other labelled samples (eg a panel of references samples).\begin{wrapfigure}{r}{0.5\textwidth}
  \begin{center}
   \includegraphics[width=0.48\textwidth]{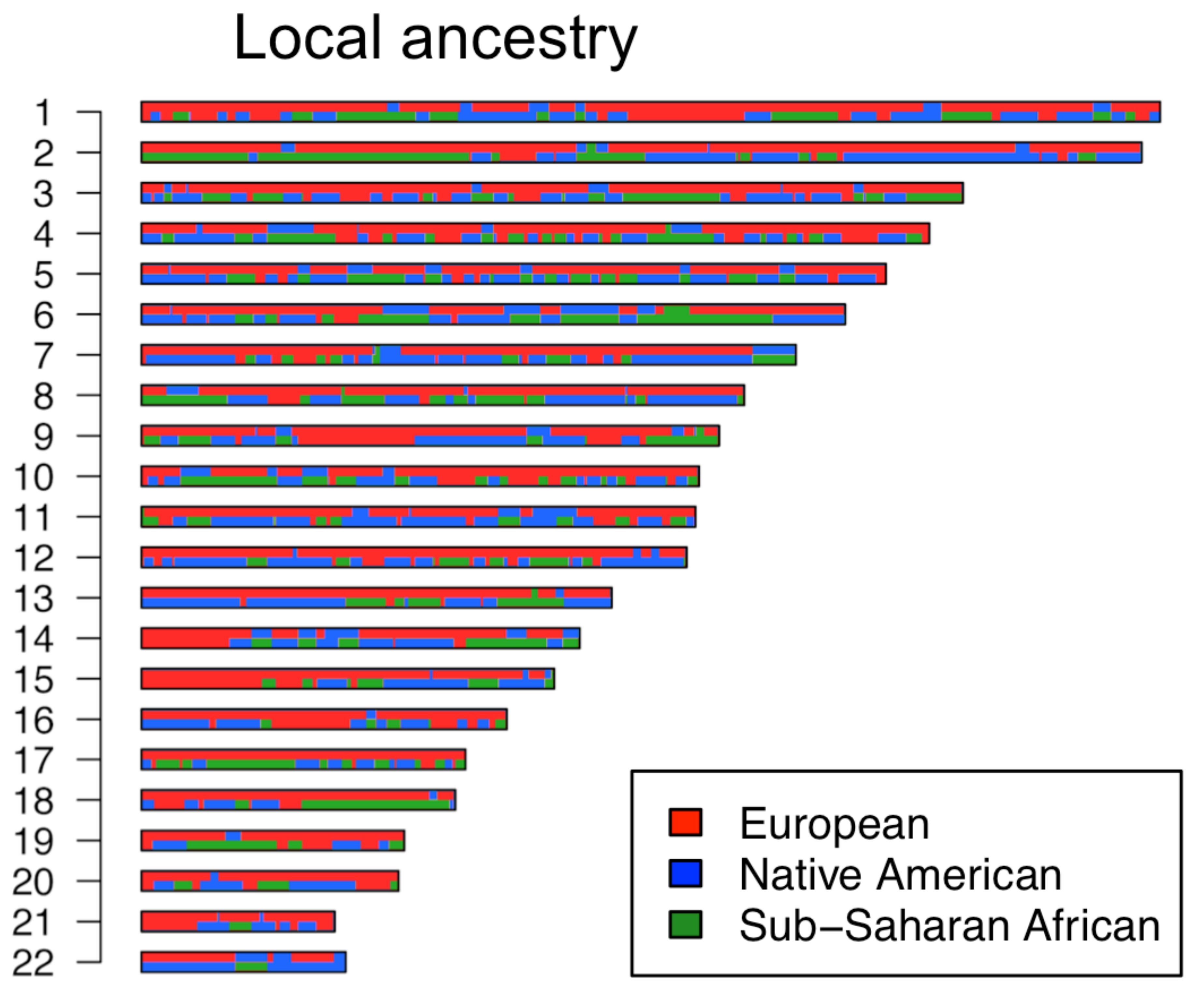}
  \end{center}
  \caption{{\small Figure from \citet{moreno2013reconstructing} the genome of a Caribbean-descent individual recruited in South Florida, USA. The 22 automosomes are coloured by ``Continental-level local ancestry calls'' \citep{moreno2013reconstructing} using an updated version of PCAdmix \citep{brisbin2012pcadmix}. 
  Reference samples used: YRI, CEU, and Native Americans from Mexico. Cropped from figure published under CC-BY 4.0. }
} \label{Fig:moreno_admixture_blocks}
\end{wrapfigure} Again, they are really statements about similarity: that 75\% of the genome is most similar to the genotypes of individuals from the CEU samples (CEPH European descent sample, Utah) and 25\% of their genotype looks similar to the genotypes from the YRI sample (Yoruba in Ibadan, Nigeria).  %who have the sample description “West African”. 

Various other finer grain methods and representations also exist. For example, there are approaches where blocks of an individual's genome (haplotypes) are statistically assigned as coming from some limited set of (often) pre-specified ancestries \citep{falush2003inference,price2009sensitive,maples2013rfmix}, defined by representative samples (see Figure \ref{Fig:moreno_admixture_blocks}).  Again, however, regions of the genome labelled as having African ancestry could simply be said to be more similar to YRI samples than to CEU samples.

Thus, all these statements about ancestry groups are really statements about the genetic similarity to other samples whose population descriptors we choose to propagate as our `ancestry group' labels. Samples are genetically similar because they share more ancestral paths, and so all of these approaches can be phrased in terms of hypotheses about shared genetic ancestors; framed thus, they can quite be a reasonable set of analyses and hypotheses for future investigation to learn about population history.  However, in most applications ancestry group labels are used as simple sample descriptors, which raises a large set of issues.

\subsection*{Issues arising from the use of genetic ancestry group labels }
A wide variety of issues have been repeatedly raised over the use of genetic ancestry labels. A  common and fair criticism is that the genetic ancestry descriptors used are often based on continental geographic labels and so overlap with racial labels \citep{weiss2009non,fujimura2011different,panofsky2017ambiguity,lewis2022getting}.  The ancestral populations alluded to by genetic ancestry labels are at most simple statistical modelling constructs, but they can easily become reified into a discrete view of ancestral populations. While this problem is most apparent in the deliberate misuse of genetic ancestry to reinforce racist narratives of human diversity \citep{panofsky2021white}, it also opens a number of pitfalls for researchers themselves, primarily because such labels obscure the heterogeneity within `ancestry groups' and the continuum of relatedness across them and in doing so can bias our thinking about genetic and environmental variation. For example, in GWAS individuals of `European genetic ancestry' are often grouped together for analyses across multiple countries. While there can be good methodological justifications for grouping samples, there are also implicit decisions about who shares enough similarity in genetics and environments to be grouped together. 
% jed twitter thread: https://twitter.com/JedMSP/status/922496778812968960

A number of technical issues can also be raised about the use of genetic ancestry labels. For example, even if we accept the idea of using ancestry groups based on genetic similarity, the resolution and naming of ancestry groups is a function of reference samples used. For example, by some methods discussed above a person living in the middle east might find themselves assigned as having “European” ancestry if middle Eastern samples are not included in the reference set. As a result of this, changes to the panels used to assess ancestry groups can also lead to confusing changes in ancestry labels, the most obvious setting where this occurs is in personal genomics but in practice it can also occur as datasets are reused across scientific papers.  These issues will not be resolved by including finer-scale sampling in reference panels, as in the limit, there will often be a fairly continuous spread of peoples’ genotypes, and so there is no natural place to carve human diversity to assign `ancestries'.

Another important aspect is the time frame implicitly assumed by descriptions of ancestry groups. Indeed, although statements of ancestry usually bracket genetic ancestors at a specific time period, this aspect is usually missing from the descriptions in papers.  For example, in analyses of samples in the Americas, many people will have generations of recent ancestors who also lived in the Americas. However, through the choice of ancestry reference panels the analysis of genetic ancestry is often implicitly targeted at describing the locations of the genetic ancestors of American people $>$600 years in the past. Moreover, it is now clear that there have been large-scale movements of people over the past 10,000 years, which makes labels based on current-day sampling locations complicated. For example, my genetic ancestors likely lived in Europe, the Middle East, and the Russian Steppe 10,000 years ago \citep{lazaridis2014ancient,allentoft2015population,haak2015massive,chintalapati2022spatiotemporal}. So these statements of genetic ancestry are best thought of as descriptions of genetic ancestors in a bracketed time period of 600 years ago up to a few thousand years ago. Quite why the field has settled on this time period as a basis for comparison is often unclear. 

\subsection*{Responses }
Partially in response to some of these criticisms, a number of alternative approaches have been put forward for population descriptors based on ancestry.  One idea might simply be to switch from using terms such as `European ancestry' to `European ancestries' to avoid implying that there is one homogeneous ancestral population underlying patterns of similarity. This proposal seems reasonable if we are to keep using ancestry group labels but does not move away from using broad geographic labels. A second, more substantial proposal is to move towards fine-grained ancestry labels, partially motivated by the reasonable objective of moving away from continental labels.

\begin{wrapfigure}{r}{0.31\textwidth}
  \begin{center}
   \includegraphics[width=0.3\textwidth]{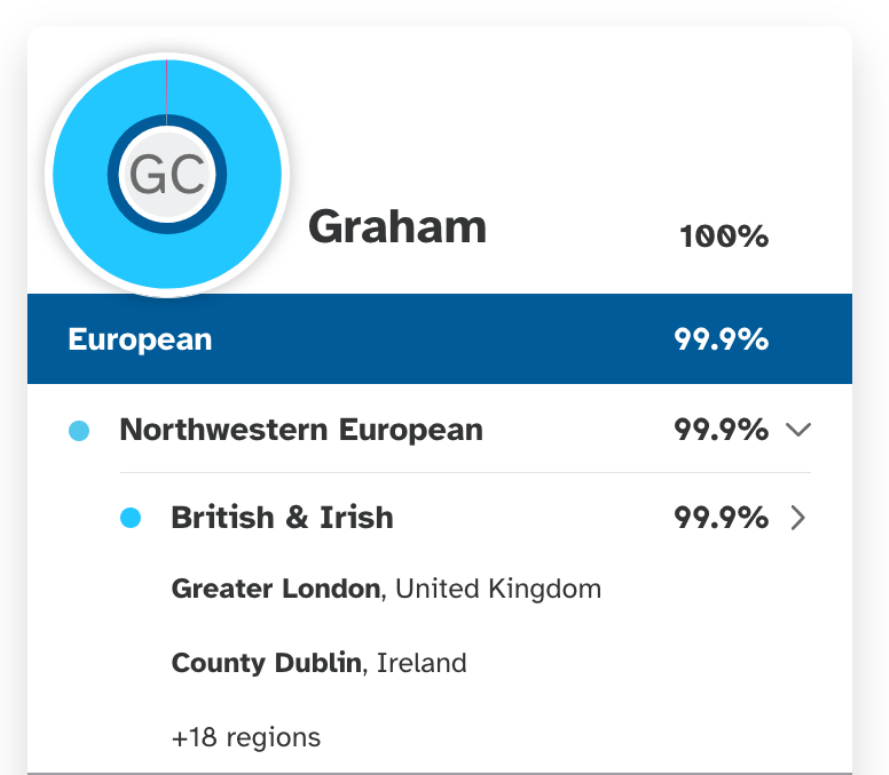}
  \end{center}
  \caption{{\small My 23andMe ancestry composition (accessed June 2022).}
} \label{Fig:Coop23}
\end{wrapfigure} For example, 23\&me  breaks down my ancestry into Northwestern European, British and Irish ancestry, and to even more fine-grained `ancestries' such as `Greater London' (Fig. \ref{Fig:Coop23}). This is based on the similarity of my haplotypes to other present-day people whose grandparents all come from these regions \citep{23andMe_ancestry,durand2021scalable}. These fine-scale approaches are also being used to examine ancestry in medical settings \citep{belbin2021toward}. However, if we have trouble defining what we mean precisely by ancestry at broad geographic scales, we will run into even more difficulties being precise about what we mean by labels such as `British and Irish' genetic ancestry, let alone `Greater London' genetic ancestry. 

Other approaches to ancestry have been laid out where a person’s ancestry could be reported for various different time epochs, which would remove the need for the implicit choice of time period. Or we could imagine tracing genetic ancestors back across geographical space over the generations, which would allow a more continuous view of ancestry \citep{osmond2021estimating,wohns2022unified}. Such approaches may well be useful for population geneticists and genetic anthropologists interested in human history. Indeed, advances in population genetics and genetic anthropology combined with ancient DNA are making large inroads into describing human history, and all these fields should work to integrate a better set of descriptors of genetic ancestors. However, the vast amount of research in human genetics (notably that funded by the NIH) is not about selling personal genomics or studying human history but rather on understanding the role of genetics in human health and disease. 

An additional justification often offered for the use of genetic ancestry is that in addition to genetics it also captures socio-environmental factors that can co-vary with ancestry. On this basis in many cases ancestry group labels are used to subset data or as a covariate in analyses to capture non-genetic effects \citep[e.g.][]{covid2021mapping}. However, in doing so, there is an issue of mixing social, geographic, and genetic labels. For example, does a set of `East Asian ancestry' polygenic scores refer to people of East Asian ancestry in the US or from the Japanese biobank? And why is East Asian the correct ancestry label for what may be a narrower subset of people? Such distinctions seem already to be important in a number of cases  \citep{giannakopoulou2021genetic}. Yet too often, studies centre the genetic ancestry label and so confound together relatedness, social environment, and sampling location. This conflation sets up a situation where it is all too easy to slip into viewing differences in genetic ancestry as a genetic cause of differences in phenotypic and health outcomes between groups. 

Along similar lines, within racial or ethnic groupings the proportion of an individual's genome from particular “ancestries” can be correlated with phenotypic outcomes due to environmental factors as well as generations of racism and discrimination. For example, in African Americans in the US, the proportion of African ancestry is correlated with geography, socioeconomic outcomes, and patterns of migration out of the American South \citep{micheletti2020genetic,baharian2016great}. But all too often papers reporting correlations with genetic ancestry in recently admixed populations do not acknowledge the potential for socio-environmental confounding and can slip straight to discussions of genetic causes.

\section*{Towards ‘genetic similarity’ descriptors}
Arguably, much of human genetics cares about matching for genetic similarity, not ancestry, when making a set of comparisons or assembling a set of controls. Consider the following commonly posed questions: “What is the disease risk of someone of this specific genotype with this broader genetic background?”; “Are my biological replicates and controls appropriately matched for this gene expression study?”; or “What set of polygenic score weights should I use for prediction for this person or haplotype?”. In all these cases, the genetics questions we are asking about are based on matching people based on genetic similarity, not a vague sense of who a person’s ancestors were. 

Researchers are also often interested in controlling for the environment, and given that in many cases the relevant environmental variables may be unknown or unmeasured, it may be reasonable to use non-genetic sample descriptors in these analyses as a control for unmeasured environmental factors. In some cases, a researcher may lack socially relevant sample descriptors and so want to use a genetic label as a proxy for a social label to account for unmeasured environmental factors. Indeed, whenever we include a genetic similarity variable (for example, a PC) in an analysis of traits, it may become a proxy for both genetic and correlated environments and social variables \citep{vilhjalmsson2013nature}. In such cases, we should be clear that correlates between trait outcomes and genetic similarity could be the result of both genetic and environmental causes rather than relying on the idea of `genetic ancestry' to telegraph that idea. 

%Given the issues associated with ‘genetic ancestry groups’, I believe that much of human genetics research should move away from using genetic ancestry labels towards more readily interpretable statements about genetic similarity (and relatedness) for sample descriptions: 

As a field we should move away from genetic ancestry labels and towards simple statements of genetic similarity: “This sample/haplotype is genetically similar to the XX sample set (in comparisons to YYY samples using ZZZ metric)” is much closer to how population genetic methods can be used to provide genetic sample descriptors. For example, “Graham is genetically similar to the GBR 1000 Genome samples (on the first 10 PC)” rather than “Graham has Northwestern European genetic ancestry”. The former sounds a little more awkward, but that awkwardness reflects the truth of how these labels work and comes with many fewer built-in assumptions and pitfalls.

From a technical standpoint, moving toward using genetic similarity labels puts a focus on how similar we need to make the match and by what measure we judge similarity. It also directs attention to the panels used to judge similarity and forces us to ask ourselves whether our panels are representative and fine-grained enough for the comparison we wish to make. 

Importantly, in my view, the term `genetically similar to' also helps to avoid the assumption of homogeneity within labels; `similar to' does not imply `same as'. Similarity-based sample descriptors also move us some way to acknowledging the continuous nature of genetic variation across human groups in our sample descriptions. I am more genetically similar to some samples than I am to others, but that does not imply that there are natural groupings. Nor do similarity-based labels imply how I, as an individual, might choose to identify or what distribution of environments I might experience. For example, a person may be genetically similar to 1000 Genomes samples labeled Southern Asian, yet in itself this similarity does not identify them as Southern Asian, whereas stating that a person has `Southern Asian genetic ancestry' comes much closer to making that linkage in people’s minds. 

Working out the genealogical history of individuals and of groups of people from around the world is a fascinating area of research, but it should not be the day job of the majority of researchers in the field of human genetics, who instead need accurate sample descriptors of current day genetic diversity that aid clear communication.

\bibliographystyle{genetics}
\bibliography{NAS_response}

\begin{thebibliography}{}

\bibitem[\protect\citeauthoryear{{23andMe}}{\textsc{{23andMe}}}{2020}]{23andMe_ancestry}
\textsc{{23andMe}}, 2020\ \ Ancestry Composition: 23andMe's State-of-the-Art
  Geographic Ancestry Analysis.
\newblock \url{https://www.23andme.com/ancestry-composition-guide/}.

\bibitem[\protect\citeauthoryear{Alexander, Novembre, and
  Lange}{\textsc{Alexander} {\em et~al.\@}}{2009}]{alexander2009fast}
\textsc{Alexander, D.~H.}, \textsc{J.~Novembre}, and \textsc{K.~Lange}, 2009\ \
  Fast model-based estimation of ancestry in unrelated individuals.
\newblock Genome research~{\em 19\/}(9)\textbf{:}\ 1655--1664.

\bibitem[\protect\citeauthoryear{Allentoft, Sikora, Sj{\"o}gren, Rasmussen,
  Rasmussen, Stenderup, Damgaard, Schroeder, Ahlstr{\"o}m, Vinner, {\it
  et~al.}}{\textsc{Allentoft} {\em et~al.\@}}{2015}]{allentoft2015population}
\textsc{Allentoft, M.~E.}, \textsc{M.~Sikora}, \textsc{K.-G. Sj{\"o}gren},
  \textsc{S.~Rasmussen}, \textsc{M.~Rasmussen}, \textsc{J.~Stenderup},
  \textsc{P.~B. Damgaard}, \textsc{H.~Schroeder}, \textsc{T.~Ahlstr{\"o}m},
  \textsc{L.~Vinner}, and \textsc{others}, 2015\ \ Population genomics of
  bronze age Eurasia.
\newblock Nature~{\em 522\/}(7555)\textbf{:}\ 167--172.

\bibitem[\protect\citeauthoryear{Baharian, Barakatt, Gignoux, Shringarpure,
  Errington, Blot, Bustamante, Kenny, Williams, Aldrich, {\it
  et~al.}}{\textsc{Baharian} {\em et~al.\@}}{2016}]{baharian2016great}
\textsc{Baharian, S.}, \textsc{M.~Barakatt}, \textsc{C.~R. Gignoux},
  \textsc{S.~Shringarpure}, \textsc{J.~Errington}, \textsc{W.~J. Blot},
  \textsc{C.~D. Bustamante}, \textsc{E.~E. Kenny}, \textsc{S.~M. Williams},
  \textsc{M.~C. Aldrich}, and \textsc{others}, 2016\ \ The great migration and
  African-American genomic diversity.
\newblock PLoS genetics~{\em 12\/}(5)\textbf{:}\ e1006059.

\bibitem[\protect\citeauthoryear{Belbin, Cullina, Wenric, Soper, Glicksberg,
  Torre, Moscati, Wojcik, Shemirani, Beckmann, {\it et~al.}}{\textsc{Belbin}
  {\em et~al.\@}}{2021}]{belbin2021toward}
\textsc{Belbin, G.~M.}, \textsc{S.~Cullina}, \textsc{S.~Wenric}, \textsc{E.~R.
  Soper}, \textsc{B.~S. Glicksberg}, \textsc{D.~Torre}, \textsc{A.~Moscati},
  \textsc{G.~L. Wojcik}, \textsc{R.~Shemirani}, \textsc{N.~D. Beckmann}, and
  \textsc{others}, 2021\ \ Toward a fine-scale population health monitoring
  system.
\newblock Cell~{\em 184\/}(8)\textbf{:}\ 2068--2083.

\bibitem[\protect\citeauthoryear{Biddanda, Rice, and
  Novembre}{\textsc{Biddanda} {\em et~al.\@}}{2020}]{biddanda2020variant}
\textsc{Biddanda, A.}, \textsc{D.~P. Rice}, and \textsc{J.~Novembre}, 2020\ \ A
  variant-centric perspective on geographic patterns of human allele frequency
  variation.
\newblock Elife~{\bf 9}\textbf{:}\ e60107.

\bibitem[\protect\citeauthoryear{Brisbin, Bryc, Byrnes, Zakharia, Omberg,
  Degenhardt, Reynolds, Ostrer, Mezey, and Bustamante}{\textsc{Brisbin} {\em
  et~al.\@}}{2012}]{brisbin2012pcadmix}
\textsc{Brisbin, A.}, \textsc{K.~Bryc}, \textsc{J.~Byrnes},
  \textsc{F.~Zakharia}, \textsc{L.~Omberg}, \textsc{J.~Degenhardt},
  \textsc{A.~Reynolds}, \textsc{H.~Ostrer}, \textsc{J.~G. Mezey}, and
  \textsc{C.~D. Bustamante}, 2012\ \ PCAdmix: principal components-based
  assignment of ancestry along each chromosome in individuals with admixed
  ancestry from two or more populations.
\newblock Human biology~{\em 84\/}(4)\textbf{:}\ 343.

\bibitem[\protect\citeauthoryear{Brown, Ye, Price, Zaitlen, {Asian Genetic
  Epidemiology Network Type 2 Diabetes Consortium}, {\it
  et~al.}}{\textsc{Brown} {\em et~al.\@}}{2016}]{brown2016transethnic}
\textsc{Brown, B.~C.}, \textsc{C.~J. Ye}, \textsc{A.~L. Price},
  \textsc{N.~Zaitlen}, \textsc{{Asian Genetic Epidemiology Network Type 2
  Diabetes Consortium}}, and \textsc{others}, 2016\ \ Transethnic
  genetic-correlation estimates from summary statistics.
\newblock The American Journal of Human Genetics~{\em 99\/}(1)\textbf{:}\
  76--88.

\bibitem[\protect\citeauthoryear{Bycroft, Fernandez-Rozadilla, Ruiz-Ponte,
  Quintela, Carracedo, Donnelly, and Myers}{\textsc{Bycroft} {\em
  et~al.\@}}{2019}]{bycroft2019patterns}
\textsc{Bycroft, C.}, \textsc{C.~Fernandez-Rozadilla}, \textsc{C.~Ruiz-Ponte},
  \textsc{I.~Quintela}, \textsc{{\'A}.~Carracedo}, \textsc{P.~Donnelly}, and
  \textsc{S.~Myers}, 2019\ \ Patterns of genetic differentiation and the
  footprints of historical migrations in the Iberian Peninsula.
\newblock Nature communications~{\em 10\/}(1)\textbf{:}\ 1--14.

\bibitem[\protect\citeauthoryear{Bycroft, Freeman, Petkova, Band, Elliott,
  Sharp, Motyer, Vukcevic, Delaneau, O’Connell, {\it
  et~al.}}{\textsc{Bycroft} {\em et~al.\@}}{2018}]{bycroft2018uk}
\textsc{Bycroft, C.}, \textsc{C.~Freeman}, \textsc{D.~Petkova},
  \textsc{G.~Band}, \textsc{L.~T. Elliott}, \textsc{K.~Sharp},
  \textsc{A.~Motyer}, \textsc{D.~Vukcevic}, \textsc{O.~Delaneau},
  \textsc{J.~O’Connell}, and \textsc{others}, 2018\ \ The UK Biobank resource
  with deep phenotyping and genomic data.
\newblock Nature~{\em 562\/}(7726)\textbf{:}\ 203--209.

\bibitem[\protect\citeauthoryear{Byeon, Islamaj, Yeganova, Wilbur, Lu, Brody,
  and Bonham}{\textsc{Byeon} {\em et~al.\@}}{2021}]{byeon2021evolving}
\textsc{Byeon, Y. J.~J.}, \textsc{R.~Islamaj}, \textsc{L.~Yeganova},
  \textsc{W.~J. Wilbur}, \textsc{Z.~Lu}, \textsc{L.~C. Brody}, and
  \textsc{V.~L. Bonham}, 2021\ \ Evolving use of ancestry, ethnicity, and race
  in genetics research—A survey spanning seven decades.
\newblock The American Journal of Human Genetics~{\em 108\/}(12)\textbf{:}\
  2215--2223.

\bibitem[\protect\citeauthoryear{Byrne, van Rheenen, van~den Berg, Veldink, and
  McLaughlin}{\textsc{Byrne} {\em et~al.\@}}{2020}]{byrne2020dutch}
\textsc{Byrne, R.~P.}, \textsc{W.~van Rheenen}, \textsc{L.~H. van~den Berg},
  \textsc{J.~H. Veldink}, and \textsc{R.~L. McLaughlin}, 2020\ \ Dutch
  population structure across space, time and GWAS design.
\newblock Nature communications~{\em 11\/}(1)\textbf{:}\ 1--11.

\bibitem[\protect\citeauthoryear{Chintalapati, Patterson, and
  Moorjani}{\textsc{Chintalapati} {\em
  et~al.\@}}{2022}]{chintalapati2022spatiotemporal}
\textsc{Chintalapati, M.}, \textsc{N.~Patterson}, and \textsc{P.~Moorjani},
  2022\ \ The spatiotemporal patterns of major human admixture events during
  the European Holocene.
\newblock eLife~{\bf 11}\textbf{:}\ e77625.

\bibitem[\protect\citeauthoryear{Coop}{\textsc{Coop}}{2013}]{Coop_How_many_genetic_anc}
\textsc{Coop, G.}, 2013\ \ How many genetic ancestors do I have?
\newblock
  \url{https://gcbias.org/2013/11/11/how-does-your-number-of-genetic-ancestors-grow-back-over-time/}.

\bibitem[\protect\citeauthoryear{Coop}{\textsc{Coop}}{2017a}]{Coop_Genetic_ancs}
\textsc{Coop, G.}, 2017a\ \ Where did your genetic ancestors come from?
\newblock \url{https://gcbias.org/2017/12/19/1628/}.

\bibitem[\protect\citeauthoryear{Coop}{\textsc{Coop}}{2017b}]{Coop_genealogical}
\textsc{Coop, G.}, 2017b\ \ Your ancestors lived all over the world.
\newblock
  \url{https://gcbias.org/2017/11/28/your-ancestors-lived-all-over-the-world/}.

\bibitem[\protect\citeauthoryear{{COVID-19 Host Genetics
  Initiative}}{\textsc{{COVID-19 Host Genetics
  Initiative}}}{2021}]{covid2021mapping}
\textsc{{COVID-19 Host Genetics Initiative}}, 2021\ \ Mapping the human genetic
  architecture of COVID-19.
\newblock Nature~{\em 600\/}(7889)\textbf{:}\ 472--477.

\bibitem[\protect\citeauthoryear{Donnelly}{\textsc{Donnelly}}{1983}]{donnelly1983probability}
\textsc{Donnelly, K.~P.}, 1983\ \ The probability that related individuals
  share some section of genome identical by descent.
\newblock Theoretical population biology~{\em 23\/}(1)\textbf{:}\ 34--63.

\bibitem[\protect\citeauthoryear{Durand, Do, Wilton, Mountain, Auton, Poznik,
  and Macpherson}{\textsc{Durand} {\em et~al.\@}}{2021}]{durand2021scalable}
\textsc{Durand, E.~Y.}, \textsc{C.~B. Do}, \textsc{P.~R. Wilton}, \textsc{J.~L.
  Mountain}, \textsc{A.~Auton}, \textsc{G.~D. Poznik}, and \textsc{J.~M.
  Macpherson}, 2021\ \ A scalable pipeline for local ancestry inference using
  tens of thousands of reference haplotypes.
\newblock bioRxiv.

\bibitem[\protect\citeauthoryear{Excoffier, Dupanloup, Huerta-S{\'a}nchez,
  Sousa, and Foll}{\textsc{Excoffier} {\em
  et~al.\@}}{2013}]{excoffier2013robust}
\textsc{Excoffier, L.}, \textsc{I.~Dupanloup}, \textsc{E.~Huerta-S{\'a}nchez},
  \textsc{V.~C. Sousa}, and \textsc{M.~Foll}, 2013\ \ Robust demographic
  inference from genomic and SNP data.
\newblock PLoS genetics~{\em 9\/}(10)\textbf{:}\ e1003905.

\bibitem[\protect\citeauthoryear{Falush, Stephens, and
  Pritchard}{\textsc{Falush} {\em et~al.\@}}{2003}]{falush2003inference}
\textsc{Falush, D.}, \textsc{M.~Stephens}, and \textsc{J.~K. Pritchard}, 2003\
  \ Inference of population structure using multilocus genotype data: linked
  loci and correlated allele frequencies.
\newblock Genetics~{\em 164\/}(4)\textbf{:}\ 1567--1587.

\bibitem[\protect\citeauthoryear{Fan, Hansen, Lo, and Tishkoff}{\textsc{Fan}
  {\em et~al.\@}}{2016}]{fan2016going}
\textsc{Fan, S.}, \textsc{M.~E. Hansen}, \textsc{Y.~Lo}, and \textsc{S.~A.
  Tishkoff}, 2016\ \ Going global by adapting local: A review of recent human
  adaptation.
\newblock Science~{\em 354\/}(6308)\textbf{:}\ 54--59.

\bibitem[\protect\citeauthoryear{Fujimura and Rajagopalan}{\textsc{Fujimura}
  and \textsc{Rajagopalan}}{2011}]{fujimura2011different}
\textsc{Fujimura, J.~H.} and \textsc{R.~Rajagopalan}, 2011\ \ Different
  differences: The use of ‘genetic ancestry’ versus race in biomedical
  human genetic research.
\newblock Social Studies of Science~{\em 41\/}(1)\textbf{:}\ 5--30.

\bibitem[\protect\citeauthoryear{Giannakopoulou, Lin, Meng, Su, Kuo, Peterson,
  Awasthi, Moscati, Coleman, Bass, {\it et~al.}}{\textsc{Giannakopoulou} {\em
  et~al.\@}}{2021}]{giannakopoulou2021genetic}
\textsc{Giannakopoulou, O.}, \textsc{K.~Lin}, \textsc{X.~Meng}, \textsc{M.-H.
  Su}, \textsc{P.-H. Kuo}, \textsc{R.~E. Peterson}, \textsc{S.~Awasthi},
  \textsc{A.~Moscati}, \textsc{J.~R. Coleman}, \textsc{N.~Bass}, and
  \textsc{others}, 2021\ \ The genetic architecture of depression in
  individuals of East Asian ancestry: a genome-wide association study.
\newblock JAMA psychiatry~{\em 78\/}(11)\textbf{:}\ 1258--1269.

\bibitem[\protect\citeauthoryear{Gravel, Henn, Gutenkunst, Indap, Marth, Clark,
  Yu, Gibbs, Project, Bustamante, {\it et~al.}}{\textsc{Gravel} {\em
  et~al.\@}}{2011}]{gravel2011demographic}
\textsc{Gravel, S.}, \textsc{B.~M. Henn}, \textsc{R.~N. Gutenkunst},
  \textsc{A.~R. Indap}, \textsc{G.~T. Marth}, \textsc{A.~G. Clark},
  \textsc{F.~Yu}, \textsc{R.~A. Gibbs}, \textsc{.~G. Project}, \textsc{C.~D.
  Bustamante}, and \textsc{others}, 2011\ \ Demographic history and rare allele
  sharing among human populations.
\newblock Proceedings of the National Academy of Sciences~{\em
  108\/}(29)\textbf{:}\ 11983--11988.

\bibitem[\protect\citeauthoryear{Gutenkunst, Hernandez, Williamson, and
  Bustamante}{\textsc{Gutenkunst} {\em
  et~al.\@}}{2009}]{gutenkunst2009inferring}
\textsc{Gutenkunst, R.~N.}, \textsc{R.~D. Hernandez}, \textsc{S.~H.
  Williamson}, and \textsc{C.~D. Bustamante}, 2009\ \ Inferring the joint
  demographic history of multiple populations from multidimensional SNP
  frequency data.
\newblock PLoS genetics~{\em 5\/}(10)\textbf{:}\ e1000695.

\bibitem[\protect\citeauthoryear{Haak, Lazaridis, Patterson, Rohland, Mallick,
  Llamas, Brandt, Nordenfelt, Harney, Stewardson, {\it et~al.}}{\textsc{Haak}
  {\em et~al.\@}}{2015}]{haak2015massive}
\textsc{Haak, W.}, \textsc{I.~Lazaridis}, \textsc{N.~Patterson},
  \textsc{N.~Rohland}, \textsc{S.~Mallick}, \textsc{B.~Llamas},
  \textsc{G.~Brandt}, \textsc{S.~Nordenfelt}, \textsc{E.~Harney},
  \textsc{K.~Stewardson}, and \textsc{others}, 2015\ \ Massive migration from
  the steppe was a source for Indo-European languages in Europe.
\newblock Nature~{\em 522\/}(7555)\textbf{:}\ 207--211.

\bibitem[\protect\citeauthoryear{Han, Carbonetto, Curtis, Wang, Granka, Byrnes,
  Noto, Kermany, Myres, Barber, {\it et~al.}}{\textsc{Han} {\em
  et~al.\@}}{2017}]{han2017clustering}
\textsc{Han, E.}, \textsc{P.~Carbonetto}, \textsc{R.~E. Curtis},
  \textsc{Y.~Wang}, \textsc{J.~M. Granka}, \textsc{J.~Byrnes},
  \textsc{K.~Noto}, \textsc{A.~R. Kermany}, \textsc{N.~M. Myres}, \textsc{M.~J.
  Barber}, and \textsc{others}, 2017\ \ Clustering of 770,000 genomes reveals
  post-colonial population structure of North America.
\newblock Nature communications~{\em 8\/}(1)\textbf{:}\ 1--12.

\bibitem[\protect\citeauthoryear{Haworth, Mitchell, Corbin, Wade, Dudding,
  Budu-Aggrey, Carslake, Hemani, Paternoster, Smith, {\it
  et~al.}}{\textsc{Haworth} {\em et~al.\@}}{2019}]{haworth2019apparent}
\textsc{Haworth, S.}, \textsc{R.~Mitchell}, \textsc{L.~Corbin}, \textsc{K.~H.
  Wade}, \textsc{T.~Dudding}, \textsc{A.~Budu-Aggrey}, \textsc{D.~Carslake},
  \textsc{G.~Hemani}, \textsc{L.~Paternoster}, \textsc{G.~D. Smith}, and
  \textsc{others}, 2019\ \ Apparent latent structure within the UK Biobank
  sample has implications for epidemiological analysis.
\newblock Nature communications~{\em 10\/}(1)\textbf{:}\ 1--9.

\bibitem[\protect\citeauthoryear{Hudson}{\textsc{Hudson}}{1990}]{hudson1990gene}
\textsc{Hudson, R.~R.}, 1990\ \ Gene genealogies and the coalescent process.
\newblock Oxford surveys in evolutionary biology~{\em 7\/}(1)\textbf{:}\ 44.

\bibitem[\protect\citeauthoryear{Jeong and Di~Rienzo}{\textsc{Jeong} and
  \textsc{Di~Rienzo}}{2014}]{jeong2014adaptations}
\textsc{Jeong, C.} and \textsc{A.~Di~Rienzo}, 2014\ \ Adaptations to local
  environments in modern human populations.
\newblock Current opinion in genetics \& development~{\bf 29}\textbf{:}\ 1--8.

\bibitem[\protect\citeauthoryear{Kelleher, Wong, Wohns, Fadil, Albers, and
  McVean}{\textsc{Kelleher} {\em et~al.\@}}{2019}]{kelleher2019inferring}
\textsc{Kelleher, J.}, \textsc{Y.~Wong}, \textsc{A.~W. Wohns},
  \textsc{C.~Fadil}, \textsc{P.~K. Albers}, and \textsc{G.~McVean}, 2019\ \
  Inferring whole-genome histories in large population datasets.
\newblock Nature Genetics~{\em 51\/}(9)\textbf{:}\ 1330--1338.

\bibitem[\protect\citeauthoryear{Lazaridis, Patterson, Mittnik, Renaud,
  Mallick, Kirsanow, Sudmant, Schraiber, Castellano, Lipson, {\it
  et~al.}}{\textsc{Lazaridis} {\em et~al.\@}}{2014}]{lazaridis2014ancient}
\textsc{Lazaridis, I.}, \textsc{N.~Patterson}, \textsc{A.~Mittnik},
  \textsc{G.~Renaud}, \textsc{S.~Mallick}, \textsc{K.~Kirsanow}, \textsc{P.~H.
  Sudmant}, \textsc{J.~G. Schraiber}, \textsc{S.~Castellano},
  \textsc{M.~Lipson}, and \textsc{others}, 2014\ \ Ancient human genomes
  suggest three ancestral populations for present-day Europeans.
\newblock Nature~{\em 513\/}(7518)\textbf{:}\ 409--413.

\bibitem[\protect\citeauthoryear{Leslie, Winney, Hellenthal, Davison,
  Boumertit, Day, Hutnik, Royrvik, Cunliffe, Lawson, {\it
  et~al.}}{\textsc{Leslie} {\em et~al.\@}}{2015}]{leslie2015fine}
\textsc{Leslie, S.}, \textsc{B.~Winney}, \textsc{G.~Hellenthal},
  \textsc{D.~Davison}, \textsc{A.~Boumertit}, \textsc{T.~Day},
  \textsc{K.~Hutnik}, \textsc{E.~C. Royrvik}, \textsc{B.~Cunliffe},
  \textsc{D.~J. Lawson}, and \textsc{others}, 2015\ \ The fine-scale genetic
  structure of the British population.
\newblock Nature~{\em 519\/}(7543)\textbf{:}\ 309--314.

\bibitem[\protect\citeauthoryear{Lewis, Molina, Appelbaum, Dauda, Di~Rienzo,
  Fuentes, Fullerton, Garrison, Ghosh, Hammonds, {\it et~al.}}{\textsc{Lewis}
  {\em et~al.\@}}{2022}]{lewis2022getting}
\textsc{Lewis, A.~C.}, \textsc{S.~J. Molina}, \textsc{P.~S. Appelbaum},
  \textsc{B.~Dauda}, \textsc{A.~Di~Rienzo}, \textsc{A.~Fuentes}, \textsc{S.~M.
  Fullerton}, \textsc{N.~Garrison}, \textsc{N.~Ghosh}, \textsc{E.~M. Hammonds},
  and \textsc{others}, 2022\ \ Getting genetic ancestry right for science and
  society.
\newblock Science~{\em 376\/}(6590)\textbf{:}\ 250--252.

\bibitem[\protect\citeauthoryear{Lewontin}{\textsc{Lewontin}}{1972}]{Lewontin1972}
\textsc{Lewontin, R.~C.}, 1972\ \ {The Apportionment of Human Diversity}.
\newblock In T.~Dobzhansky, M.~K. Hecht, and W.~C. Steere (Eds.), {\em
  Evolutionary Biology}, Chapter~14, pp.\  381--398. New York:
  Appleton-Century-Crofts.

\bibitem[\protect\citeauthoryear{Liu, Huang, Chen, Zhao, Yuan, Francis, Fang,
  Li, Lin, Liu, {\it et~al.}}{\textsc{Liu} {\em
  et~al.\@}}{2018}]{liu2018genomic}
\textsc{Liu, S.}, \textsc{S.~Huang}, \textsc{F.~Chen}, \textsc{L.~Zhao},
  \textsc{Y.~Yuan}, \textsc{S.~S. Francis}, \textsc{L.~Fang}, \textsc{Z.~Li},
  \textsc{L.~Lin}, \textsc{R.~Liu}, and \textsc{others}, 2018\ \ Genomic
  analyses from non-invasive prenatal testing reveal genetic associations,
  patterns of viral infections, and Chinese population history.
\newblock Cell~{\em 175\/}(2)\textbf{:}\ 347--359.

\bibitem[\protect\citeauthoryear{Liu, Mao, Krause, and Fu}{\textsc{Liu} {\em
  et~al.\@}}{2021}]{liu2021insights}
\textsc{Liu, Y.}, \textsc{X.~Mao}, \textsc{J.~Krause}, and \textsc{Q.~Fu},
  2021\ \ Insights into human history from the first decade of ancient human
  genomics.
\newblock Science~{\em 373\/}(6562)\textbf{:}\ 1479--1484.

\bibitem[\protect\citeauthoryear{Manrubia, Derrida, and
  Zanette}{\textsc{Manrubia} {\em et~al.\@}}{2003}]{manrubia2003genealogy}
\textsc{Manrubia, S.~C.}, \textsc{B.~H. Derrida}, and \textsc{D.~H. Zanette},
  2003\ \ Genealogy in the Era of Genomics.
\newblock American Scientist~{\em 91\/}(2)\textbf{:}\ 158--165.

\bibitem[\protect\citeauthoryear{Mao, Bigham, Mei, Gutierrez, Weiss, Brutsaert,
  Leon-Velarde, Moore, Vargas, McKeigue, {\it et~al.}}{\textsc{Mao} {\em
  et~al.\@}}{2007}]{mao2007genomewide}
\textsc{Mao, X.}, \textsc{A.~W. Bigham}, \textsc{R.~Mei},
  \textsc{G.~Gutierrez}, \textsc{K.~M. Weiss}, \textsc{T.~D. Brutsaert},
  \textsc{F.~Leon-Velarde}, \textsc{L.~G. Moore}, \textsc{E.~Vargas},
  \textsc{P.~M. McKeigue}, and \textsc{others}, 2007\ \ A genomewide admixture
  mapping panel for Hispanic/Latino populations.
\newblock The American Journal of Human Genetics~{\em 80\/}(6)\textbf{:}\
  1171--1178.

\bibitem[\protect\citeauthoryear{Maples, Gravel, Kenny, and
  Bustamante}{\textsc{Maples} {\em et~al.\@}}{2013}]{maples2013rfmix}
\textsc{Maples, B.~K.}, \textsc{S.~Gravel}, \textsc{E.~E. Kenny}, and
  \textsc{C.~D. Bustamante}, 2013\ \ RFMix: a discriminative modeling approach
  for rapid and robust local-ancestry inference.
\newblock The American Journal of Human Genetics~{\em 93\/}(2)\textbf{:}\
  278--288.

\bibitem[\protect\citeauthoryear{Marjoram and Griffiths}{\textsc{Marjoram} and
  \textsc{Griffiths}}{1995}]{marjoram1995ancestral}
\textsc{Marjoram, P.} and \textsc{R.~Griffiths}, 1995\ \ {\em Progress in
  Population Genetics and Human Evolution, IMA Volumes in Mathematics and its
  Applications.}, Chapter An ancestral recombination graph, pp.\  257--270.

\bibitem[\protect\citeauthoryear{Martin, Gignoux, Walters, Wojcik, Gravel,
  Daly, Bustamante, and Kenny}{\textsc{Martin} {\em
  et~al.\@}}{2016}]{martin2016population}
\textsc{Martin, A.~R.}, \textsc{C.~R. Gignoux}, \textsc{R.~K. Walters},
  \textsc{G.~L. Wojcik}, \textsc{S.~Gravel}, \textsc{M.~J. Daly}, \textsc{C.~D.
  Bustamante}, and \textsc{E.~E. Kenny}, 2016\ \ Population genetic history and
  polygenic risk biases in 1000 Genomes populations.
\newblock bioRxiv\textbf{:}\ 070797.

\bibitem[\protect\citeauthoryear{Mathieson and Scally}{\textsc{Mathieson} and
  \textsc{Scally}}{2020}]{mathieson2020ancestry}
\textsc{Mathieson, I.} and \textsc{A.~Scally}, 2020\ \ What is ancestry?
\newblock PLoS Genetics~{\em 16\/}(3)\textbf{:}\ e1008624.

\bibitem[\protect\citeauthoryear{McVean}{\textsc{McVean}}{2009}]{mcvean2009genealogical}
\textsc{McVean, G.}, 2009\ \ A genealogical interpretation of principal
  components analysis.
\newblock PLoS genetics~{\em 5\/}(10)\textbf{:}\ e1000686.

\bibitem[\protect\citeauthoryear{Micheletti, Bryc, Esselmann, Freyman, Moreno,
  Poznik, Shastri, Agee, Aslibekyan, Auton, {\it et~al.}}{\textsc{Micheletti}
  {\em et~al.\@}}{2020}]{micheletti2020genetic}
\textsc{Micheletti, S.~J.}, \textsc{K.~Bryc}, \textsc{S.~G.~A. Esselmann},
  \textsc{W.~A. Freyman}, \textsc{M.~E. Moreno}, \textsc{G.~D. Poznik},
  \textsc{A.~J. Shastri}, \textsc{M.~Agee}, \textsc{S.~Aslibekyan},
  \textsc{A.~Auton}, and \textsc{others}, 2020\ \ Genetic consequences of the
  transatlantic slave trade in the Americas.
\newblock The American Journal of Human Genetics~{\em 107\/}(2)\textbf{:}\
  265--277.

\bibitem[\protect\citeauthoryear{Moreno-Estrada, Gravel, Zakharia, McCauley,
  Byrnes, Gignoux, Ortiz-Tello, Mart{\'\i}nez, Hedges, Morris, {\it
  et~al.}}{\textsc{Moreno-Estrada} {\em
  et~al.\@}}{2013}]{moreno2013reconstructing}
\textsc{Moreno-Estrada, A.}, \textsc{S.~Gravel}, \textsc{F.~Zakharia},
  \textsc{J.~L. McCauley}, \textsc{J.~K. Byrnes}, \textsc{C.~R. Gignoux},
  \textsc{P.~A. Ortiz-Tello}, \textsc{R.~J. Mart{\'\i}nez}, \textsc{D.~J.
  Hedges}, \textsc{R.~W. Morris}, and \textsc{others}, 2013\ \ Reconstructing
  the population genetic history of the Caribbean.
\newblock PLoS genetics~{\em 9\/}(11)\textbf{:}\ e1003925.

\bibitem[\protect\citeauthoryear{Narasimhan, Patterson, Moorjani, Rohland,
  Bernardos, Mallick, Lazaridis, Nakatsuka, Olalde, Lipson, {\it
  et~al.}}{\textsc{Narasimhan} {\em et~al.\@}}{2019}]{narasimhan2019formation}
\textsc{Narasimhan, V.~M.}, \textsc{N.~Patterson}, \textsc{P.~Moorjani},
  \textsc{N.~Rohland}, \textsc{R.~Bernardos}, \textsc{S.~Mallick},
  \textsc{I.~Lazaridis}, \textsc{N.~Nakatsuka}, \textsc{I.~Olalde},
  \textsc{M.~Lipson}, and \textsc{others}, 2019\ \ The formation of human
  populations in South and Central Asia.
\newblock Science~{\em 365\/}(6457)\textbf{:}\ eaat7487.

\bibitem[\protect\citeauthoryear{{National Academies of Sciences, Engineering,
  and Medicine}}{\textsc{{National Academies of Sciences, Engineering, and
  Medicine}}}{2022}]{NAS_descriptors}
\textsc{{National Academies of Sciences, Engineering, and Medicine}}, 2022\ \
  Use of Race, Ethnicity, and Ancestry as Population Descriptors in Genomics
  Research.
\newblock
  \url{https://www.nationalacademies.org/our-work/use-of-race-ethnicity-and-ancestry-as-population-descriptors-in-genomics-research}.

\bibitem[\protect\citeauthoryear{Nelson, Wegmann, Ehm, Kessner, St.~Jean,
  Verzilli, Shen, Tang, Bacanu, Fraser, {\it et~al.}}{\textsc{Nelson} {\em
  et~al.\@}}{2012}]{nelson2012abundance}
\textsc{Nelson, M.~R.}, \textsc{D.~Wegmann}, \textsc{M.~G. Ehm},
  \textsc{D.~Kessner}, \textsc{P.~St.~Jean}, \textsc{C.~Verzilli},
  \textsc{J.~Shen}, \textsc{Z.~Tang}, \textsc{S.-A. Bacanu},
  \textsc{D.~Fraser}, and \textsc{others}, 2012\ \ An abundance of rare
  functional variants in 202 drug target genes sequenced in 14,002 people.
\newblock Science~{\em 337\/}(6090)\textbf{:}\ 100--104.

\bibitem[\protect\citeauthoryear{Novembre and Peter}{\textsc{Novembre} and
  \textsc{Peter}}{2016}]{novembre2016recent}
\textsc{Novembre, J.} and \textsc{B.~M. Peter}, 2016\ \ Recent advances in the
  study of fine-scale population structure in humans.
\newblock Current opinion in genetics \& development~{\bf 41}\textbf{:}\
  98--105.

\bibitem[\protect\citeauthoryear{Olalde, Brace, Allentoft, Armit, Kristiansen,
  Booth, Rohland, Mallick, Sz{\'e}cs{\'e}nyi-Nagy, Mittnik, {\it
  et~al.}}{\textsc{Olalde} {\em et~al.\@}}{2018}]{olalde2018beaker}
\textsc{Olalde, I.}, \textsc{S.~Brace}, \textsc{M.~E. Allentoft},
  \textsc{I.~Armit}, \textsc{K.~Kristiansen}, \textsc{T.~Booth},
  \textsc{N.~Rohland}, \textsc{S.~Mallick}, \textsc{A.~Sz{\'e}cs{\'e}nyi-Nagy},
  \textsc{A.~Mittnik}, and \textsc{others}, 2018\ \ The Beaker phenomenon and
  the genomic transformation of northwest Europe.
\newblock Nature~{\em 555\/}(7695)\textbf{:}\ 190--196.

\bibitem[\protect\citeauthoryear{Osmond and Coop}{\textsc{Osmond} and
  \textsc{Coop}}{2021}]{osmond2021estimating}
\textsc{Osmond, M.~M.} and \textsc{G.~Coop}, 2021\ \ Estimating dispersal rates
  and locating genetic ancestors with genome-wide genealogies.
\newblock bioRxiv.

\bibitem[\protect\citeauthoryear{Panofsky and Bliss}{\textsc{Panofsky} and
  \textsc{Bliss}}{2017}]{panofsky2017ambiguity}
\textsc{Panofsky, A.} and \textsc{C.~Bliss}, 2017\ \ Ambiguity and scientific
  authority: population classification in genomic science.
\newblock American Sociological Review~{\em 82\/}(1)\textbf{:}\ 59--87.

\bibitem[\protect\citeauthoryear{Panofsky, Dasgupta, and
  Iturriaga}{\textsc{Panofsky} {\em et~al.\@}}{2021}]{panofsky2021white}
\textsc{Panofsky, A.}, \textsc{K.~Dasgupta}, and \textsc{N.~Iturriaga}, 2021\ \
  How White nationalists mobilize genetics: From genetic ancestry and human
  biodiversity to counterscience and metapolitics.
\newblock American Journal of Physical Anthropology~{\em 175\/}(2)\textbf{:}\
  387--398.

\bibitem[\protect\citeauthoryear{Patel, Musharoff, Spence, Pimentel,
  Tcheandjieu, Mostafavi, Sinnott-Armstrong, Clarke, Smith, Program, {\it
  et~al.}}{\textsc{Patel} {\em et~al.\@}}{2022}]{patel2022genetic}
\textsc{Patel, R.~A.}, \textsc{S.~A. Musharoff}, \textsc{J.~P. Spence},
  \textsc{H.~Pimentel}, \textsc{C.~Tcheandjieu}, \textsc{H.~Mostafavi},
  \textsc{N.~Sinnott-Armstrong}, \textsc{S.~L. Clarke}, \textsc{C.~J. Smith},
  \textsc{V.~M.~V. Program}, and \textsc{others}, 2022\ \ Genetic interactions
  drive heterogeneity in causal variant effect sizes for gene expression and
  complex traits.
\newblock The American Journal of Human Genetics~{\em 109\/}(7)\textbf{:}\
  1286--1297.

\bibitem[\protect\citeauthoryear{Patterson, Isakov, Booth, B{\"u}ster, Fischer,
  Olalde, Ringbauer, Akbari, Cheronet, Bleasdale, {\it
  et~al.}}{\textsc{Patterson} {\em et~al.\@}}{2022}]{patterson2022large}
\textsc{Patterson, N.}, \textsc{M.~Isakov}, \textsc{T.~Booth},
  \textsc{L.~B{\"u}ster}, \textsc{C.-E. Fischer}, \textsc{I.~Olalde},
  \textsc{H.~Ringbauer}, \textsc{A.~Akbari}, \textsc{O.~Cheronet},
  \textsc{M.~Bleasdale}, and \textsc{others}, 2022\ \ Large-scale migration
  into Britain during the Middle to Late Bronze Age.
\newblock Nature~{\em 601\/}(7894)\textbf{:}\ 588--594.

\bibitem[\protect\citeauthoryear{Peter, Petkova, and Novembre}{\textsc{Peter}
  {\em et~al.\@}}{2020}]{peter2020genetic}
\textsc{Peter, B.~M.}, \textsc{D.~Petkova}, and \textsc{J.~Novembre}, 2020\ \
  Genetic landscapes reveal how human genetic diversity aligns with geography.
\newblock Molecular biology and evolution~{\em 37\/}(4)\textbf{:}\ 943--951.

\bibitem[\protect\citeauthoryear{Peterson, Kuchenbaecker, Walters, Chen,
  Popejoy, Periyasamy, Lam, Iyegbe, Strawbridge, Brick, {\it
  et~al.}}{\textsc{Peterson} {\em et~al.\@}}{2019}]{peterson2019genome}
\textsc{Peterson, R.~E.}, \textsc{K.~Kuchenbaecker}, \textsc{R.~K. Walters},
  \textsc{C.-Y. Chen}, \textsc{A.~B. Popejoy}, \textsc{S.~Periyasamy},
  \textsc{M.~Lam}, \textsc{C.~Iyegbe}, \textsc{R.~J. Strawbridge},
  \textsc{L.~Brick}, and \textsc{others}, 2019\ \ Genome-wide association
  studies in ancestrally diverse populations: opportunities, methods, pitfalls,
  and recommendations.
\newblock Cell~{\em 179\/}(3)\textbf{:}\ 589--603.

\bibitem[\protect\citeauthoryear{Popejoy, Crooks, Fullerton, Hindorff, Hooker,
  Koenig, Pino, Ramos, Ritter, Wand, {\it et~al.}}{\textsc{Popejoy} {\em
  et~al.\@}}{2020}]{popejoy2020clinical}
\textsc{Popejoy, A.~B.}, \textsc{K.~R. Crooks}, \textsc{S.~M. Fullerton},
  \textsc{L.~A. Hindorff}, \textsc{G.~W. Hooker}, \textsc{B.~A. Koenig},
  \textsc{N.~Pino}, \textsc{E.~M. Ramos}, \textsc{D.~I. Ritter},
  \textsc{H.~Wand}, and \textsc{others}, 2020\ \ Clinical genetics lacks
  standard definitions and protocols for the collection and use of diversity
  measures.
\newblock The American Journal of Human Genetics~{\em 107\/}(1)\textbf{:}\
  72--82.

\bibitem[\protect\citeauthoryear{Price, Tandon, Patterson, Barnes, Rafaels,
  Ruczinski, Beaty, Mathias, Reich, and Myers}{\textsc{Price} {\em
  et~al.\@}}{2009}]{price2009sensitive}
\textsc{Price, A.~L.}, \textsc{A.~Tandon}, \textsc{N.~Patterson}, \textsc{K.~C.
  Barnes}, \textsc{N.~Rafaels}, \textsc{I.~Ruczinski}, \textsc{T.~H. Beaty},
  \textsc{R.~Mathias}, \textsc{D.~Reich}, and \textsc{S.~Myers}, 2009\ \
  Sensitive detection of chromosomal segments of distinct ancestry in admixed
  populations.
\newblock PLoS genetics~{\em 5\/}(6)\textbf{:}\ e1000519.

\bibitem[\protect\citeauthoryear{Pritchard, Stephens, and
  Donnelly}{\textsc{Pritchard} {\em et~al.\@}}{2000}]{pritchard2000inference}
\textsc{Pritchard, J.~K.}, \textsc{M.~Stephens}, and \textsc{P.~Donnelly},
  2000\ \ Inference of population structure using multilocus genotype data.
\newblock Genetics~{\em 155\/}(2)\textbf{:}\ 945--959.

\bibitem[\protect\citeauthoryear{Privé, Aschard, Carmi, Folkersen, Hoggart,
  O’Reilly, and Vilhjálmsson}{\textsc{Privé} {\em
  et~al.\@}}{2022}]{Prive2022}
\textsc{Privé, F.}, \textsc{H.~Aschard}, \textsc{S.~Carmi},
  \textsc{L.~Folkersen}, \textsc{C.~Hoggart}, \textsc{P.~F. O’Reilly}, and
  \textsc{B.~J. Vilhjálmsson}, 2022\ \ Portability of 245 polygenic scores
  when derived from the UK Biobank and applied to 9 ancestry groups from the
  same cohort.
\newblock The American Journal of Human Genetics~{\em 109\/}(1)\textbf{:}\
  12--23.

\bibitem[\protect\citeauthoryear{Ralph and Coop}{\textsc{Ralph} and
  \textsc{Coop}}{2010}]{ralph2010parallel}
\textsc{Ralph, P.} and \textsc{G.~Coop}, 2010\ \ Parallel adaptation: one or
  many waves of advance of an advantageous allele?
\newblock Genetics~{\em 186\/}(2)\textbf{:}\ 647--668.

\bibitem[\protect\citeauthoryear{Ramachandran, Deshpande, Roseman, Rosenberg,
  Feldman, and Cavalli-Sforza}{\textsc{Ramachandran} {\em
  et~al.\@}}{2005}]{ramachandran2005support}
\textsc{Ramachandran, S.}, \textsc{O.~Deshpande}, \textsc{C.~C. Roseman},
  \textsc{N.~A. Rosenberg}, \textsc{M.~W. Feldman}, and \textsc{L.~L.
  Cavalli-Sforza}, 2005\ \ Support from the relationship of genetic and
  geographic distance in human populations for a serial founder effect
  originating in Africa.
\newblock Proceedings of the National Academy of Sciences~{\em
  102\/}(44)\textbf{:}\ 15942--15947.

\bibitem[\protect\citeauthoryear{Raveane, Aneli, Montinaro, Athanasiadis,
  Barlera, Birolo, Boncoraglio, Di~Blasio, Di~Gaetano, Pagani, {\it
  et~al.}}{\textsc{Raveane} {\em et~al.\@}}{2019}]{raveane2019population}
\textsc{Raveane, A.}, \textsc{S.~Aneli}, \textsc{F.~Montinaro},
  \textsc{G.~Athanasiadis}, \textsc{S.~Barlera}, \textsc{G.~Birolo},
  \textsc{G.~Boncoraglio}, \textsc{A.~M. Di~Blasio}, \textsc{C.~Di~Gaetano},
  \textsc{L.~Pagani}, and \textsc{others}, 2019\ \ Population structure of
  modern-day Italians reveals patterns of ancient and archaic ancestries in
  Southern Europe.
\newblock Science Advances~{\em 5\/}(9)\textbf{:}\ eaaw3492.

\bibitem[\protect\citeauthoryear{Rohde, Olson, and Chang}{\textsc{Rohde} {\em
  et~al.\@}}{2004}]{rohde2004modelling}
\textsc{Rohde, D.~L.}, \textsc{S.~Olson}, and \textsc{J.~T. Chang}, 2004\ \
  Modelling the recent common ancestry of all living humans.
\newblock Nature~{\em 431\/}(7008)\textbf{:}\ 562--566.

\bibitem[\protect\citeauthoryear{Rosenberg and Nordborg}{\textsc{Rosenberg} and
  \textsc{Nordborg}}{2002}]{rosenberg2002genealogical}
\textsc{Rosenberg, N.~A.} and \textsc{M.~Nordborg}, 2002\ \ Genealogical trees,
  coalescent theory and the analysis of genetic polymorphisms.
\newblock Nature Reviews Genetics~{\em 3\/}(5)\textbf{:}\ 380--390.

\bibitem[\protect\citeauthoryear{Skoglund and Mathieson}{\textsc{Skoglund} and
  \textsc{Mathieson}}{2018}]{skoglund2018ancient}
\textsc{Skoglund, P.} and \textsc{I.~Mathieson}, 2018\ \ Ancient genomics of
  modern humans: the first decade.
\newblock Annu. Rev. Genomics Hum. Genet~{\bf 19}\textbf{:}\ 381--404.

\bibitem[\protect\citeauthoryear{Speidel, Forest, Shi, and
  Myers}{\textsc{Speidel} {\em et~al.\@}}{2019}]{speidel2019method}
\textsc{Speidel, L.}, \textsc{M.~Forest}, \textsc{S.~Shi}, and \textsc{S.~R.
  Myers}, 2019\ \ A method for genome-wide genealogy estimation for thousands
  of samples.
\newblock Nature Genetics~{\em 51\/}(9)\textbf{:}\ 1321--1329.

\bibitem[\protect\citeauthoryear{{Use of Race, Ethnicity, and Ancestry as
  Population Descriptors in Genomics Research Meetings}}{\textsc{{Use of Race,
  Ethnicity, and Ancestry as Population Descriptors in Genomics Research
  Meetings}}}{2022a}]{NAS_descriptors_1}
\textsc{{Use of Race, Ethnicity, and Ancestry as Population Descriptors in
  Genomics Research Meetings}}, 2022a\ \ Public Workshop 1.
\newblock
  \url{https://www.nationalacademies.org/event/02-14-2022/committee-on-use-of-race-ethnicity-and-ancestry-as-population-descriptors-in-genomics-research-meeting-1}.

\bibitem[\protect\citeauthoryear{{Use of Race, Ethnicity, and Ancestry as
  Population Descriptors in Genomics Research Meetings}}{\textsc{{Use of Race,
  Ethnicity, and Ancestry as Population Descriptors in Genomics Research
  Meetings}}}{2022b}]{NAS_descriptors_2}
\textsc{{Use of Race, Ethnicity, and Ancestry as Population Descriptors in
  Genomics Research Meetings}}, 2022b\ \ Public Workshop 2.
\newblock
  \url{https://www.nationalacademies.org/event/04-04-2022/committee-on-use-of-race-ethnicity-and-ancestry-as-population-descriptors-in-genomics-research-meeting-2-and-public-workshop}.

\bibitem[\protect\citeauthoryear{{Use of Race, Ethnicity, and Ancestry as
  Population Descriptors in Genomics Research Meetings}}{\textsc{{Use of Race,
  Ethnicity, and Ancestry as Population Descriptors in Genomics Research
  Meetings}}}{2022c}]{NAS_descriptors_3}
\textsc{{Use of Race, Ethnicity, and Ancestry as Population Descriptors in
  Genomics Research Meetings}}, 2022c\ \ Public Workshop 3.
\newblock
  \url{https://www.nationalacademies.org/event/06-14-2022/use-of-race-ethnicity-and-ancestry-as-population-descriptors-in-genomics-research-meeting-3-and-public-workshop}.

\bibitem[\protect\citeauthoryear{Vilhj{\'a}lmsson and
  Nordborg}{\textsc{Vilhj{\'a}lmsson} and
  \textsc{Nordborg}}{2013}]{vilhjalmsson2013nature}
\textsc{Vilhj{\'a}lmsson, B.~J.} and \textsc{M.~Nordborg}, 2013\ \ The nature
  of confounding in genome-wide association studies.
\newblock Nature Reviews Genetics~{\em 14\/}(1)\textbf{:}\ 1--2.

\bibitem[\protect\citeauthoryear{Wang, Z{\"o}llner, and
  Rosenberg}{\textsc{Wang} {\em et~al.\@}}{2012}]{wang2012quantitative}
\textsc{Wang, C.}, \textsc{S.~Z{\"o}llner}, and \textsc{N.~A. Rosenberg}, 2012\
  \ A quantitative comparison of the similarity between genes and geography in
  worldwide human populations.

\bibitem[\protect\citeauthoryear{Weiss and Long}{\textsc{Weiss} and
  \textsc{Long}}{2009}]{weiss2009non}
\textsc{Weiss, K.~M.} and \textsc{J.~C. Long}, 2009\ \ Non-Darwinian
  estimation: my ancestors, my genes' ancestors.
\newblock Genome Research~{\em 19\/}(5)\textbf{:}\ 703--710.

\bibitem[\protect\citeauthoryear{Wohns, Wong, Jeffery, Akbari, Mallick,
  Pinhasi, Patterson, Reich, Kelleher, and McVean}{\textsc{Wohns} {\em
  et~al.\@}}{2022}]{wohns2022unified}
\textsc{Wohns, A.~W.}, \textsc{Y.~Wong}, \textsc{B.~Jeffery},
  \textsc{A.~Akbari}, \textsc{S.~Mallick}, \textsc{R.~Pinhasi},
  \textsc{N.~Patterson}, \textsc{D.~Reich}, \textsc{J.~Kelleher}, and
  \textsc{G.~McVean}, 2022\ \ A unified genealogy of modern and ancient
  genomes.
\newblock Science~{\em 375\/}(6583)\textbf{:}\ eabi8264.

\bibitem[\protect\citeauthoryear{Zaidi and Mathieson}{\textsc{Zaidi} and
  \textsc{Mathieson}}{2020}]{zaidi2020demographic}
\textsc{Zaidi, A.~A.} and \textsc{I.~Mathieson}, 2020\ \ Demographic history
  mediates the effect of stratification on polygenic scores.
\newblock Elife~{\bf 9}\textbf{:}\ e61548.

\end{thebibliography}

\subsubsection*{Acknowledgments}
Thanks to Vince Buffalo, Doc Edge, Jeff Groh, Emily Josephs, James Kitchens, Magnus Nordborg, Peter Ralph, Alexis Simon, and Silu Wang for comments on a earlier draft. Funding was provided by the National Institutes of Health (NIH R35 GM136290 awarded to GC). The genesis of this piece was a talk given at the \citeauthor{NAS_descriptors} “Use of Race, Ethnicity, and Ancestry as Population Descriptors in Genomics Research” workshop, I thank the organizers for the opportunity to organize my thoughts on this topic.

\end{document}